\documentclass[12pt,preprint]{emulateapj}

\shorttitle{Sulfurization of Iron in the Dynamic Solar Nebula}
\shortauthors{Fred J. Ciesla}

\begin{document}


\title{Sulfurization of Iron in the Dynamic Solar Nebula and Implications for Planetary Compositions}

\author{Fred J. Ciesla\altaffilmark{1}}
\affil{Department of the Geophysical Sciences, The University of Chicago, 5734 South Ellis Avenue, Chicago, IL 60637}
\newpage

\begin{abstract}

One explanation for the enhanced ratio of volatiles to hydrogen in Jupiter's atmosphere compared to a a gas of solar composition is that the planet accreted volatile-bearing clathrates during its formation.  Models, however, suggest that S would be over abundant if clathrates were the primary carrier of Jupiter's volatiles.  This led to the suggestion that S was depleted in the outer nebula due to the formation troilite (FeS).  Here, this depletion is quantitatively explored by modeling the coupled dynamical and chemical evolution of Fe grains in the solar nebula.  It is found that disks that undergo rapid radial expansion from an initially compact state may allow sufficient production of FeS and carry H$_{2}$S-depleted gas outward where ices would form, providing the conditions needed for S-depleted clathrates to form.  However, this expansion would also carry FeS grains to this region, which could also be incorporated into planetesimals.  Thus for clathrates to be a viable source of volatiles, models must account for the presence of both H$_{2}$S in FeS in the outer solar nebula.

\end{abstract}

\keywords{planets and satellites: formation -- planets and satellites: composition -- planets and satellites: individual (Jupiter) -- protoplanetary disks}

\newpage

\section{Introduction}

 In the core accretion model, Jupiter's atmosphere is largely derived from solar nebula gas that was gravitationally captured by a $\sim$10 Earth mass solid core, with additional contribution from planetesimals that were accreted during the planets formation \citep{pollack96,hubickyj05}.  As the nebular gas would have had roughly a solar composition, any deviations in the composition of the planet's atmosphere from that of the Sun is thought to reflect the contribution of the late accreted planetesimals.  The \emph{Galileo} probe, during its decent, found that the
the ratios Ar/H, Kr/H, Xe/H, C/H, N/H, and S/H were  2-4 times greater than a gas of solar composition gas \citep{owen99}.   The high abundances of these volatile elements, particularly the noble gases, is puzzling as they do not react with solids and do not freeze out as pure species, and thus are not expected in planetesimals until very low temperatures ($< \sim$10 K) are reached.
 
Two mechanisms that attribute the enhanced abundances of these elements to accretion of cold, icy planetesimals have been proposed.  The first was that these elements were trapped in amorphous water ice at very low temperatures ($<$30 K) \citep{owen99}.  At these low temperatures, experiments have shown that gaseous molecules and atoms that are adsorbed on a solid surface become trapped as water freezes out on that solid \citep{barnun85, notesco03,yokochi12}.  As very low temperatures are required to produce amorphous ice and to allow for efficient trapping, this mechanism requires that Jupiter formed from or accreted planetesimals from much further away from the Sun than its current location, or that temperatures in the Jupiter-formation zone were much lower than predicted by current models for the solar nebula.  

The second mechanism involves forming clathrate hydrates, crystalline ice with guest molecules trapped inside \citep{luninestevenson85,gautier01}.    \citet{gautier01} calculated the abundances of volatiles that would be trapped as clathrates formed in a cooling solar nebula and found that the predicted abundances of noble gases, C, and N would match the abundances inferred from the \emph{Galileo} probe measurements.  However, they found that S would be present at a greater abundance than was observed.

To explain the discrepancy in the predicted and observed S abundances, \citet{gautier01} suggested that H$_{2}$S, the primary gaseous carrier of S in the planet formation region was depleted due to the formation of troilite via the reaction:
$Fe_{(s)} + H_{2}S_{(g)} \rightarrow FeS_{(s)} + H_{2 (g)}$.  Troilite is commonly found in meteoritic materials, and the reaction occurs on timescales that are short compared to the lifetime of the solar nebula \citep{fegley88,lauretta96, tachibana98}. \citet{gautier01} argued that only $\sim$35\% of H$_{2}$S would need to be removed via this reaction to match the \emph{Galileo} probe measurements, and suggested that as FeS formed in the terrestrial planet region of the solar nebula, H$_{2}$S depleted gas diffused out to the Jupiter formation region, allowing for lower amounts of S to be locked up in clathrates than would otherwise occur.  While some use such a depletion for a starting point in giant planet formation models \citep[e.g.][]{mousis14}, \citet{atreya03} challenged this idea, calling the removal of this much H$_{2}$S  ``quantitatively unsupportable.''

\citet{lodders04} offered an alternative explanation for the volatile abundance.  Citing the low oxygen abundance measured by the probe,  \citet{lodders04} suggested Jupiter formed inside the snow line.  All volatiles were then accreted directly from the solar nebula gas, with additional C and N delivered in the form of organics.  In this model sulfur could trace the contribution of solids and gas accreted by the planet: if S was distributed purely in the gas, then its relative abundance should be the same as all other gaseous species; if S was present as a solid (FeS) then its abundance would reflect the amount of solids accreted by Jupiter as it formed.  \citet{lodders04} noted that the S-to-noble gas ratio was solar, and assumed that it was delivered as FeS, and thus concluded that Jupiter formed from a solar mix of solids and nebular gas.

The behavior and speciation of S in the solar nebula and its distribution between the gas and solids is thus critical to understanding how volatiles were delivered to a young Jupiter.
\citet{pasek05}  explored this issue by modeling the evolving concentration of H$_{2}$S within a turbulent, diffusive protoplanetary disk. H$_{2}$S was assumed to be removed from the gas due to FeS formation where temperatures fell in the range of 500-690 K.  The upper bound of 690 K was chosen as this was where FeS  became stable in the nebula, while the lower bound of 500 K was set because reaction kinetics were too slow to produce much FeS.  Any H$_{2}$S that diffused into the region of the nebula that fell in this temperature range was assumed to immediately react with Fe and the sulfur permanently locked up there as FeS.  As a result, the H$_{2}$S abundance decreased throughout the nebula over time, reaching $<$65\% of its initial value inside $\sim$6 AU in 10$^{5}$ years, as predicted by \citet{gautier01}.

While providing important insights into the H$_{2}$S evolution in the solar nebula,
the treatment of FeS formation was somewhat simplified in \citet{pasek05}.  For example, chemical kinetics were calculated assuming the pressure and temperature conditions at the disk midplane and a solar abundance of H$_{2}$S.  Due to diffusion, grains would have been mixed throughout the vertical extent of the disk,  where temperatures and pressures were much lower, and reaction timescales longer \citep{ciesla10}.  Further, the lower abundances of H$_{2}$S as the disk evolved implies longer reaction times as the rate of FeS formation is directly proportional to the partial pressure of H$_{2}$S.  Finally, particles would have migrated in and out of the FeS formation region on timescales less than or comparable to the chemical timescales, suggesting that complete reaction between H$_{2}$S and Fe would not necessarily occur.  


Here a new model that tracks the motions of dust grains in a dynamically evolving protoplanetary disk is used to expand the study of \citet{pasek05}  and quantify in detail the extent to which Fe and S would react prior to Jupiter's formation, with  the focus on  the level to which H$_{2}$S becomes depleted as a result of troilite formation and the implications for the S abundance where clathrates could form.  The modeling methods are described in Section 2.  In Section 3,  the results for two examples of protoplanetary disk evolutionary histories are presented.  These results and the implications for the determining factors for planetary compositions in our Solar System are discussed at the conclusion of the paper.

\section{Model}

\subsection{Particle and Disk Evolution}

Particle tracking methods \citep{ciesla10,ciesla11,cieslasandford12,charnoz11} are employed to determine the exact physical environments the Fe/FeS grains were exposed to within the evolving solar nebula.  The evolution of the protoplanetary disk is modeled using the classical $\alpha$-viscosity formalism, where stresses within the differentially rotating gas drive mass and angular momentum transport \citep{ss73,lyndenbell74,hartmann98}, which allows us to consider the evolution of the disk over $\sim$10$^{6}$ years of evolution.    The viscosity in the disk is given by $\nu$=$\alpha c_{s} H$, where $c_{s}$ is the local speed of sound, $H$ is the local disk scale-height, and $\alpha$ is a parameter describing the strength of the viscosity.  The motions of the grains in the disk were calculated accounting for vertical settling by gravity, gas drag migration, advective flows associated with disk evolution, and diffusion, where the diffusivity was taken to equal to the disk viscosity ($\mathcal{D} \sim \nu$).  A similar model was used by \citet{pasek05}, though only diffusive redistribution of the gas was considered; the disk structure was not allowed to evolve during the time of interest.

Iron particles of a given radius, $a$, were suspended across the disk at constant Fe/H$_{2}$ mass ratio of $\epsilon$=1.67$\times$10$^{-3}$ \citep{palme14}.  The mass of the particles was set at $m_{grain}$=$\frac{4}{3}\pi a^{3} \rho_{Fe}$ where $\rho_{Fe}$=7.874 g/cm$^{3}$.  As tracking the dynamics of each iron grain in the solar nebula is computationally prohibitive (a $M_{disk}$=0.1$M_{\odot}$ disk would contain $\sim$1.67$\times$10$^{-4}$ $M_{\odot}$ of Fe, which would amount to $\sim$10$^{40}$ Fe grains 1 $\mu$m grains), each model grain is treated as a \emph{super-grain} with its evolution representing the evolution of a number of other grains equal to $N_{rep} = \epsilon M_{disk}$/$\left(m_{grain} n_{part} \right)$,  where $n_{part}$ is the number of super-grains considered in the simulation.   The super-grains were distributed radially within the disk to ensure a solar-ratio of Fe/H$_{2}$ throughout.  Grains were initially placed at the disk midplane, however vertical mixing leads to grains becoming well mixed vertically on short timescales \citep{ciesla10}, making the choice of initial $z_{0}$ unimportant for our results.

\subsection{Gas-Grain Chemistry}

Iron grains  sulfidize through reaction with H$_{2}$S  to produce troilite, while troilite decomposes following the reverse reaction:
$
FeS_{(s)} + H_{2 (g)} \rightarrow Fe_{(s)} + H_{2}S_{(g)}
$.
Because of the low pressures expected in protoplanetary disks, one of the primary controls on the rate of gas-grain reactions is the collision rate of reacting gas molecules with the grains \citep{fegley88,fegley00}.  The collision rate of a gaseous species, $i$, with a grain is:
\begin{equation}
\gamma_{i}= 2.5 \times 10^{31} \frac{P_{i}}{M_{i} T} 4 \pi a^{2} 
\end{equation}
where $P_{i}$ is the pressure of the species in bars, $M_{i}$ is the molecular weight of the gaseous molecule, $T$ is the temperature of the gas and solid, and $a$ is the radius of the grain.  The fraction of such collisions that would result in the reaction proceeding would be determined by the activation energy of the reaction, $E$, and given by
%
$f = e^{-\frac{E}{R T}}$
%
where $R$ is the gas constant.  Experiments found that the activation energy for troilite formation (the forward reaction here) falls in the range of 42-50 kJ/mol \citep{pasek05,lauretta05}.  We adopt a value of $E_{f}$=50 kJ/mol, but have found lower values in this range do not change the results significantly. The reverse reaction, where the S in FeS combines with H$_{2}$ to form H$_{2}$S gas leaving Fe behind, can be determined using the equilibrium constant for the reaction as in \citet{lauretta96}, giving $E_{r}$=115 kJ/mol.  The rate at which iron is converted into FeS is given by:
\begin{equation}
\frac{d N_{Fe}}{dt} = -4 \pi a^{2} \left( \gamma_{H_{2}S}f_{f} \beta - \gamma_{H_{2}}f_{r} \left(1 - \beta \right) \right)
\end{equation}
where $N_{Fe}$ is the number of Fe atoms in a grain and $\beta$ is the fraction of iron in the grain remaining in metallic form, while $\left( 1- \beta \right)$ represented the fraction of iron taken up as FeS.   Only small, micron-sized grains are considered so that the diffusion of S in the grain is not a rate-limiting step, and thus can be ignored \citep{pasek05}.

\subsection{Calculation Method}
Two illustrative cases are considered, each with 
a 0.1 M$_{\odot}$ disk with an initial surface density profile given by $\Sigma \left( r \right)$=$\Sigma_{0} \left( \frac{r}{r_{0}} \right)^{-1}$ extending from $r$=0.05 AU to $r_{max}$, with $r_{max}$=10 and 50 AU at the beginning of the simulations.  In each case $\alpha$=10$^{-3}$.
  Grain movement is calculated using the particle tracking methods in \citet{cieslasandford12} \citep[see also][]{ciesla10,ciesla11,charnoz11}.  In carrying out the calculations, the disk evolves for a time step, $\Delta t$=$\frac{1}{6}$ yr, after which time the surface density and temperature are updated everywhere.  The movement of the grains are then determined using the local disk properties.  The chemical evolution of these grains is calculated using the conditions at the new location.  In calculating the chemical evolution of each grain, it is assumed that the solids remain in contact with the same gas throughout the course of a calculation.  That is, as troilite forms in a grain, that grain continues to react with the H$_{2}$S-depleted gas.  Grains that migrate within $r$=0.5 AU of the Sun are assumed to be lost from the disk, carrying whatever S they contain with them.  In all cases,  $n_{part}$=2x10$^{5}$ super-particles are considered in the model, with each grain representing $\sim$5$\times$10$^{34}$ actual iron grains.

The vertical structure of the disk is calculated assuming a purely viscous model as the focus is on  the early evolution of the disk when mass and angular momentum transport are rapid and thus internal heating would dominate.  High temperatures are expected during this early epoch \citep{boss98}.  At later times, mass transport may be limited to the surface layers of the disk and irradiative heating would control the thermal structure leading to temperatures throughout the disk would be so low to prevent significant reaction between H$_{2}$S and Fe or decomposition of FeS.  This meaning that iron sulfidization was limited to the very early stages of solar nebula evolution.  Note that the changes in H$_{2}$ abundance due to chemical reactions are minimal and do not impact the overall surface density of the disk anywhere.


\section{Results}
 
Figure 1 shows the results of the two cases described above, with the H$_{2}$S distribution in the gas, normalized to its solar value, shown throughout the inner 10 AU of the disk at times of 0.25, 0.5, 0.75 and 1 Myr.  The distribution is found counting the particles at each time in a series of radial bins and calculating the amount of H$_{2}$S remaining in for each particle, then dividing by the area to get the surface density of the gas.  This is then normalized to the abundance of H$_{2}$S that would be expected for a solar S/H ratio, given the local surface density.

Valleys in the H$_{2}$S distribution are readily apparent, centered in the 0.5-1.5 AU range in each case.  These valleys correspond to where FeS was readily produced through reaction of Fe with H$_{2}$S, as midplane temperatures ranged here from $\sim$400-700 K. Viscous disks begin in a very hot state due to the high rates of dissipation (heat production) and high optical depths (heat retention), but cool with time as they lose mass to their central stars and expand due to angular momentum conservation, thinning the disk and allowing radiation to escape more readily. Thus the region of FeS production migrates closer to the star with time.  Inside of  these valleys temperatures were higher leading to decomposition of FeS and the return of H$_{2}$S to the gas, while in the outer disk temperatures were too low for much FeS to form and any FeS found there, or H$_{2}$S removed, is a result of reactions in the inner disk and outward transport through the disk.  The gradual decrease in H$_{2}$S near the inner edge of the disk is an artifact of the calculations and arises due to the removal of particles from the simulation inside of 0.5 AU to represent loss by accretion onto the star.  

While similarities exist between the results presented here and those of \citet{pasek05}, in that H$_{2}$S abundances can decrease in the disk over time, key differences are apparent.  Most important among these is that the concentration of H$_{2}$S does not decrease indefinitely here as it did in \citet{pasek05}.  Instead, the amount of H$_{2}$S outside the FeS-formation valley remain relatively constant for the $r_{max}$=50 AU case.  While outward diffusion carries 
H$_{2}$S depleted outward in the disk, the net motions of material due to viscous evolution in the disk are inward in this region.  Thus the H$_{2}$S-depleted gas can only diffuse outward so far before being pushed inwards by the advective motions associated with disk evolution.    However, the initially more compact disk ($r_{max}$=10 AU)
experiences very rapid outward expansion in this region of the disk, meaning that H$_{2}$S-depleted gas not only diffuses outwards, but is also carried outward by the net motions of the disk.  This leads to a significant overall depletion of 2-10 AU region shown here.  Efficient depletion occurs only very early in the history of the disk however, as once the disk expands the net motions of the 2-10 AU region become directed inward.  This allows materials which originated at greater distances, and thus did not become depleted in H$_{2}$S, to be carried inwards.  As a result, the H$_{2}$S abundance increases with time.

\section{Discussion}

Depletion of gaseous H$_{2}$S in the solar nebula as a result of troilite formation was possible in the solar nebula.  The greatest depletion would have occurred in the inner nebula, where physical conditions allowed the reaction to occur on short timescales. Depletion levels of 65\% or greater in the outer solar nebula as needed in the clathration model of \citet{gautier01}, require that the disk begin in a massive, compact configuration.
This compact structure  would have allowed FeS to form in the inner disk with H$_{2}$S-depleted gas being carried outward during the radial expansion of the disk that occurred during its early viscous evolution.  Such a depletion is more difficult to achieve in disks which begin with a more extended structure, as the net flow for the inner disk is directed towards the star. 

An initially compact disk would result in significant outward radial transport of materials beyond just Fe and FeS grains or H$_{2}$S gas.  \citet{ciesla10} and \citet{jacquet11} argued that a similar configuration would aid in the outward transport and preservation of Calcium, Aluminum-rich Inclusions (CAIs) found in meteorites.    Indeed, this dynamical evolution could lead to significant mixing of inner solar nebula materials into the outer solar nebula as postulated by \citet{prinnfegley89} and be consistent with the high abundance of refractory materials found in the Stardust samples from Comet Wild 2 \citep{brownlee06}.  Thus this evolution may be consistent with other evidence for radial mixing and transport in the disk.

In the calculations presented here, H$_{2}$S was depleted only through reaction with Fe to produce FeS, however other reactions may occur, incorporating S into solids.  For example, H$_{2}$S may react with NH$_{3}$ gas at lower temperatures to form NH$_{4}$SH solids.  The importance of this reaction depends on the extent to which N is present as NH$_{3}$: equilibrium chemical models predict that NH$_{3}$ should be the dominant form, but sluggish kinetics may keep much of N in the form of N$_{2}$ \citep{lewisprinn80}.  The exact level of conversion will depend on the P-T conditions that gas was exposed to during solar nebula evolution.  This issue will be the focus of future work, using the methods outlined here.  

While H$_{2}$S can be depleted in the outer solar nebula, this region is not predicted to be depleted in total S.   FeS grains would be carried outward by the same dynamical evolution which carried H$_{2}$S-depleted gas to this region.  \citet{taylor04} criticized the model of \citet{gautier01} on the grounds that removal of the needed sulfur in the gas phase would lead to too much FeS in the inner solar nebula, which cannot be reconciled with the observed compositions of the terrestrial planets.  Here, we find that FeS would not remain in the inner solar nebula, but would be mixed and made available to outer Solar System bodies.  As a result, the S content of planetesimals would be higher than predicted by clathrate models alone: while a lower abundance of H$_{2}$S allows the proper S/H ratio to be achieved in the ices described by \citet{gautier01}, additional S would be found in the rocky component of the dust that these planetesimals formed from. This suggests that H$_{2}$S must have been depleted to an even greater degree if clathrate-bearing plantesimals are to properly account for the compositions of Jupiter's atmosphere.  It is unlikely all S is locked up in rocky materials, as \citet{boissier07} found that H$_{2}$S emissions from Comet Hale-Bopp were consistent with the molecule being present in comet itself, and not the product of photochemistry in the coma. Additional measurements of S abundances and its speciation in comets will provide important insights into how this element was distributed and the extent to which it may have been locked up as a rocky mineral during the early, hot phase of solar nebula evolution.

\emph{Acknowledgments} The author is grateful for stimulating conversations with Matthew Pasek and Jonathan Lunine, as well as very insightful comments  from the referee which improved the paper greatly.  This work was supported by NASA grant NNX14AQ17G from the Outer Planets Research Program.

\bibliographystyle{apj}


\newpage
\begin{figure}
\begin{center}
\includegraphics[width=4.5in]{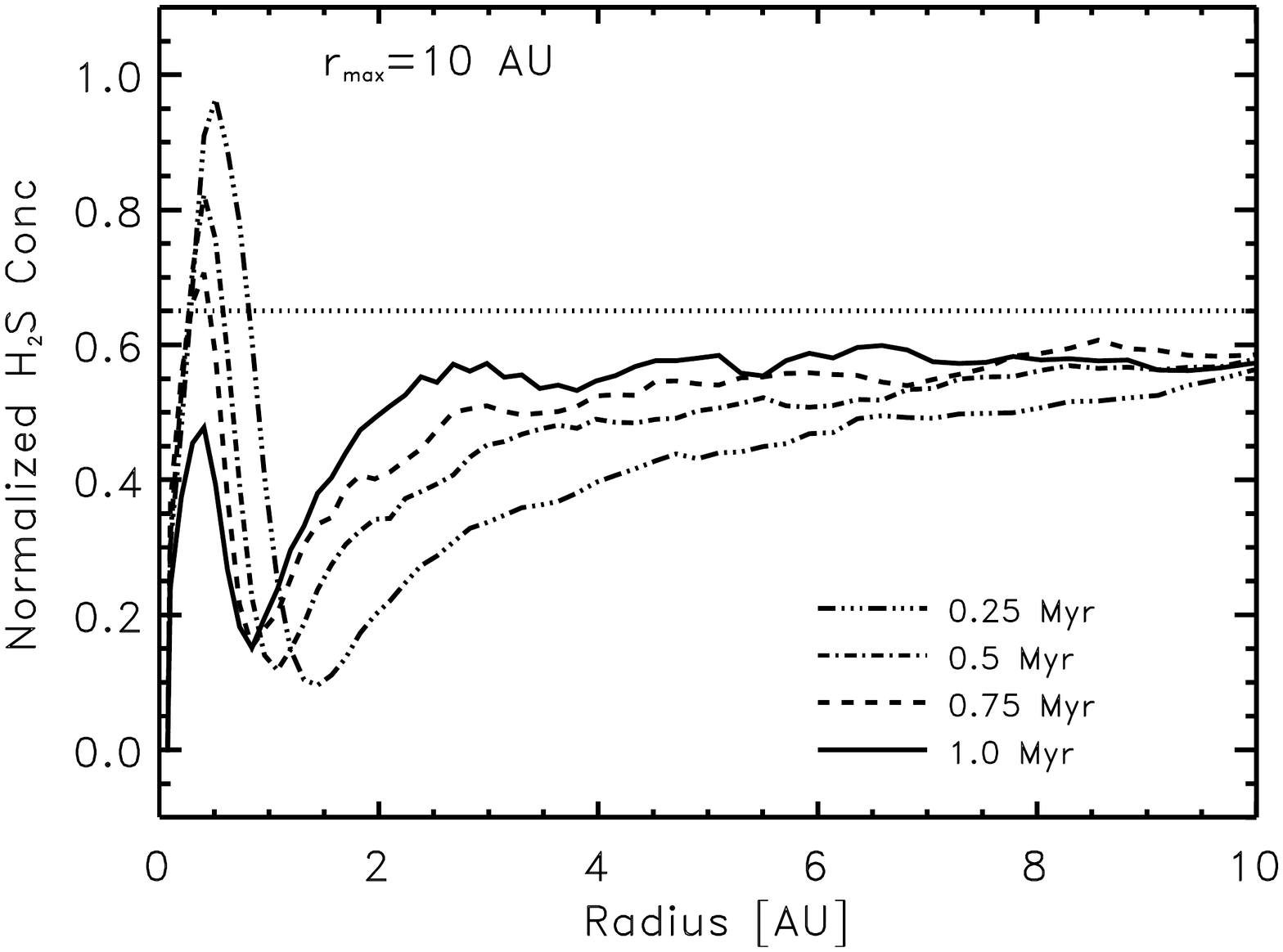}\\
\includegraphics[width=4.5in]{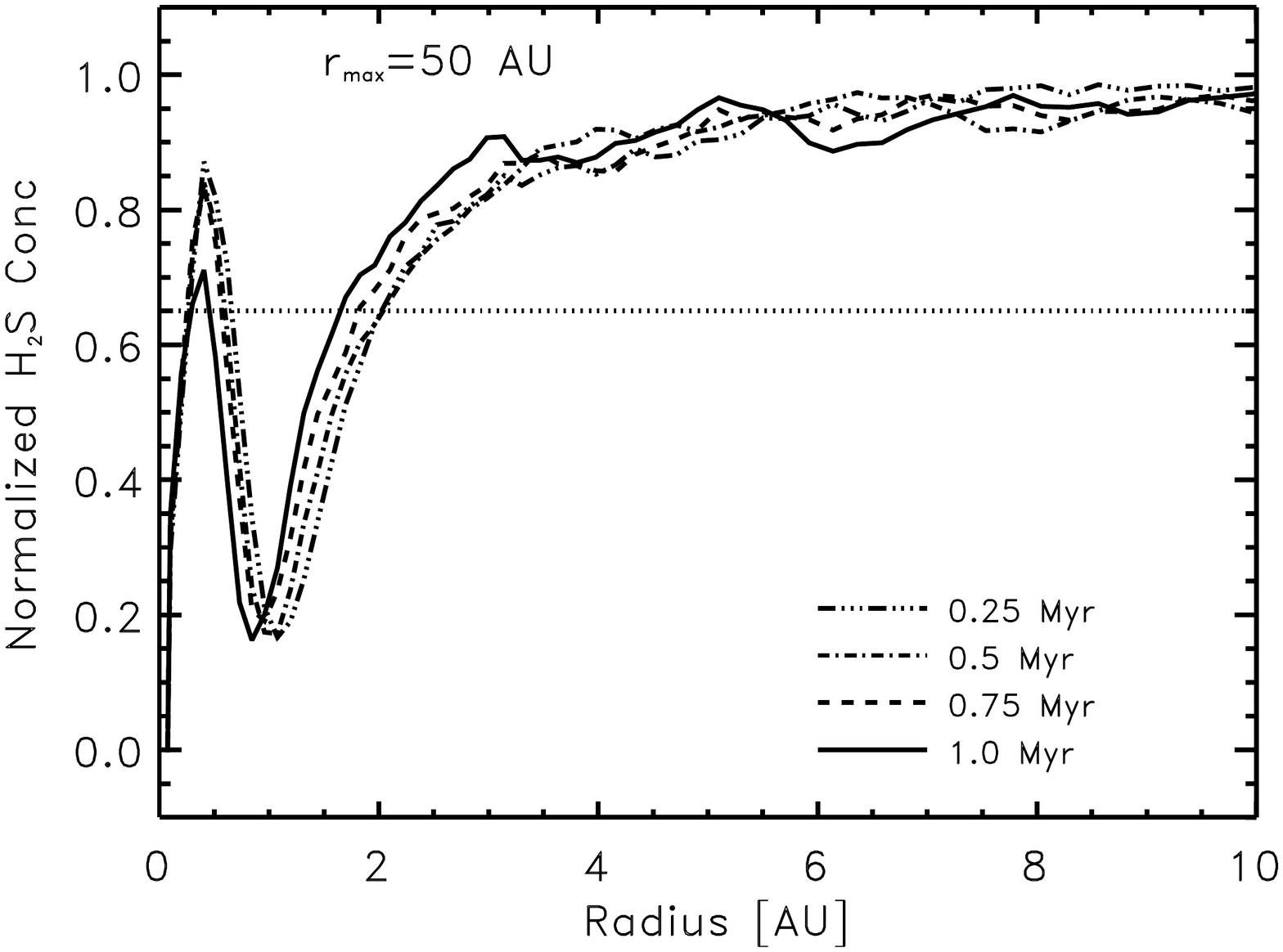}
\end{center}
\caption{Fraction of H$_{2}$S remaining in the gas phase in the inner 10 AU of the disk which begins with $r_{max}$=10 AU and 50 AU.  The distribution is shown at various times during the first 10$^{6}$ years of evolution. }
\end{figure}

\end{document}